# LIVEIA:

## A Light-based Immersive Visualization Environment for Imaginative Actualization

Liane Gabora
Department of Psychology
University of British Columbia
Kelowna, BC, Canada, V1V 1V7
liane.gabora@ubc.ca

*Abstract*—This paper describes an immersive and interactive visualization environment that uses light as a metaphor for psychological phenomena. Creative life force is portrayed as ambient light, and people's psyches are represented by spheres that amplify and transform light. Personality characteristics, situations, and relationships are systematically depicted using a systematic visual language based on the properties of light and how it interacts with physical objects. The technology enables users to visualize and creatively experiment with light-based representations of themselves and others, including patterns of interaction and how they have come about, and how they could change and unfold in the future.

*Keywords - interactive; light; metaphor; psyche; psychological model*

## I. INTRODUCTION

The association between light and creative or 'enlightened' psychological states has been woven into the human psyche since the 'dawn' of civilization [1]. It permeates our language, as in: creative *spark*, moment of *illumination*, *flash* of insight, *brilliant* idea, *bright* versus *dim*-witted, and show me the *light*. It even appears in cartoons: everyone knows what it means when a lightbulb appears above Charlie Brown's head. Religious history is replete with accounts of something not just vaguely light-like but a rarefied light that is felt rather than seen, and seems to burn from within. Eskimo shamen called it qaumaneq. Vedanta Hinduists call it Atma. The Tibetan Book of the Dead refers to it as the clear light of Buddha-nature. According to the Buddhist allegory of Indra's Net, humanity consists of a web made of threads of light stretching horizontally through space and vertically through time. At every intersection dwells an individual, and in every individual dwells a crystal bead of light. Organic processes, including cognitive processes, are made possible through the harnessing of light through photosynthesis [2], and as Carl Sagan says we are made of stardust. Thus, it is not just in a metaphorical sense that we are beings of light.

This new technology, inspired by the light metaphor, uses principles of optics [3,4,5] to guide the construction of a Light-based Immersive Visualization Environment for Imaginative Actualization (LIVEIA) that enables people to visualize and play with their inner worlds. It aims to have therapeutic value as a form of art therapy in which outputs have straightforwardly interpretable symbolic meanings. LIVEIA will derive its power not through persuasive argumentation, but through imagery that works at an intuitive gut level. Visualization has proven effective for facilitating understanding of everything from weather patterns to stock market trends, but its potential to facilitate understanding of people's inner lives and interactions, and how the future is shaped by thinking of the present, is virtually untapped.

## II. THE LOGIC UNDERLYING THE TECHNOLOGY

In LIVEIA, an individual's creative life force or chi is portrayed as ambient light, and the psyche as a sphere that amplifies and transforms light, as illustrated in **Fig. 1a**. Users are prompted to use spheres represent themselves and others with whom they interact, and instructed how to translate personality traits, areas of expertise, and unique attributes of individuals into attributes of the spheres. For example, someone with little life force is represented as having low light intensity as in **Fig. 1b**, whereas someone who is aloof, preoccupied, or who hides their true nature with a protective mask is represented with an opaque shell around their sphere that reflects and locally traps their light as in **Fig. 1c**.

Thoughts are visualized as beams of light with different waveforms. Complex thoughts are represented as superpositions of simpler waveforms. The greater the perceived importance or emotional valence of a particular thought, that stronger the intensity of the beam. The greater the extent to which a thought deviates from the concrete "here and now", i.e., the greater the extent to which it lies in the realm of the imagination, the more diffuse the beam. The experience of not being able to express an idea clearly to others is represented as the divergence of diffuse light as it passes out the sphere, as illustrated in **Fig. 2a**.

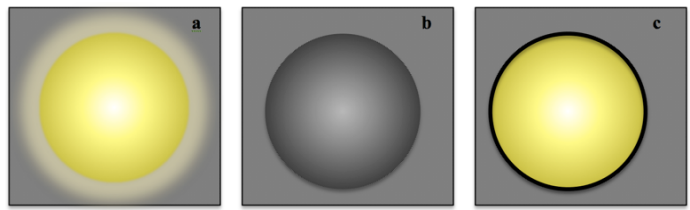

Figure 1. (a) A sphere made of a material with a higher refractive index than air (e.g., crystal) traps and amplifies light (left). A sphere may appear dark because (b) it *is* dark inside (center), or (c) its light is hidden (right).

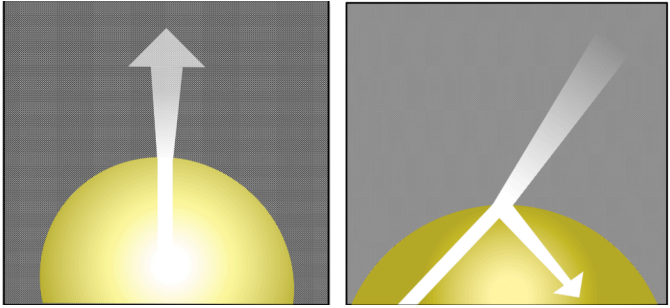

Figure 2 (a) Since the surface is concave, a ray of light diverges, or becomes less focused, after passing through the sphere. The more focused it is, the less it diverges. (b) When an incident ray (lower left) reaches the interior surface of a sphere it breaks into two: a reflected ray (which goes back into the sphere), and a refracted ray (which passes through). Since the surface is concave, the refracted ray diverges and becomes less focused as it passes through, while the reflected ray converges and becomes more focused.

The experience of *reflecting on* an idea is represented in **Fig. 2b**. The refracted ray represents the sense of focused concentration one exudes when reflecting on a problem. Since the surface is concave, the ray becomes more focused each time it reflects. This can be used to represent how, as you "bounce it around in your mind", it becomes more focused, until it achieves sufficient clarity to be articulated to others. At this point it is no longer a content of thought (unless someone brings it up or someone makes you think of it), and so no longer represented as a distinct ray but as a spark that contributes to the interior lighting of the sphere as a whole.

*Superficial or fleeting thoughts* (e.g., "I like that dress") are represented by rays originating near the inner surface of the sphere. There is only a tiny portion of the sphere through which it can project without significant distortion due to refraction, as shown in **Fig. 3a**. Everywhere else it arrives at the surface of the sphere at an angle that deviates significantly from the perpendicular. This represents that a superficial thought is are generally specific to a certain situation.

When a ray of light originates near the center it can project in any direction without distortion due to refraction because it always intersects the sphere perpendicular to its surface, as in **Fig. 3b**. Focused rays that originate close to the center represent *deep thoughts* (e.g., the concept of equality) or *deep-seated beliefs*, which are often applicable to many aspects of life. As one reflects on ideas they not only come into focus but become interwoven with more aspects of the self. This is represented as a shift in their point of origin toward the center.

It has been shown that the proclivity to deceive others is highly correlated with a distorted perception of reality [6], and this phenomenon can be understood using LIVEIA. A fracture (or vein of a different material) will cause a beam of light traveling through the sphere to bend (refract), and change direction. Deception, i.e., bending the truth, is thus represented as the deliberate use of a fracture to redirect a beam of light. When one is lying to someone else this happens at the surface of the sphere; if one is lying to oneself the refraction is occurring within the sphere.

Fracturing can represent, literally, a lack of integrity, a state wherein one is living with lies, or where one's values are not in sync with ones' actions, or one is living with memories that are too painful to face. Interestingly, the more fractured the interior, the longer it takes for a beam of light entering the sphere to reach equilibrium. Also, the greater the extent of fracturing, the less uniformly lit the interior will be when equilibrium is reached. There may be regions so fractured that light barely penetrates them. They represent the "shadow side" of the psyche, the aspects of oneself or one's life that one wants to avoid, as illustrated in **Fig. 4a**.

*Enlightenment.* When a ray of light reflects off the interior surfaces of a sphere it quickly reaches equilibrium such that it is no longer a single, distinct ray but now diffusely lights the interior. An individual who has achieved a state of enlightenment is represented as one who has no fragmentation or impurities in the psyche, such that the interior is rightly and uniformly lit, as in **Fig. 4b**. Such an individual communicates from the core of the self as opposed to a superficial level of the self, because there is nothing blocking the core. Enlightenment in this model is not a rarified state; it is a state in which one is free of internal fragmentation and able to "be themself", achievable by almost anyone. It is proposed that mindfulness is the state of remaining alert to the presence of fragmentation or shadows and considering them from different perspectives to overcome them.

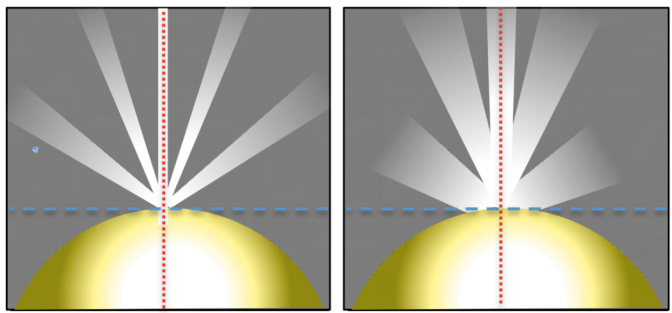

Figure 3. Left: Light originating near the periphery can only radiate in one direction; otherwise there is refraction and loss of intensity since it is not perpendicular to the sphere. Right: This effect is magnified if the beam is diffuse.

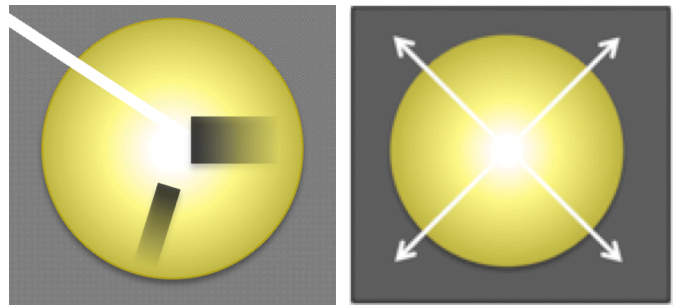

Figure 4 Left: A beam of light coming from the upper left refracts (bends and changes direction), and becomes darker and less focused as it passes through a fracture, or may be unable to penetrate the fracture entirely. A fractured sphere may not be uniformly lit; it may contain shadowed regions that the light does not fully penetrate. Right: Light coming from the center can radiate out in any direction with no refraction and minimal reflection because wherever it contacts the sphere it is perpendicular to it.

*Complex Thoughts and Feelings.* In addition to enabling the depiction of general properties of a psyche as we have seen above, the metaphor enables people to break thought patterns down into their constituent components by representing specific feelings, knowledge, values, and assumptions that make up a recurring pattern of thought as light of different colors and frequencies. Let us say, for example, that a compulsive desire to check that the door is locked is represented by the high-frequency pattern depicted in **Fig. 5,** cell **a**. One might come to realize that there is something lurking behind that compulsion: a memory of being intruded upon. This is depicted in cell **b**, giving rise to the more complex pattern in cell **c** that combines the two. The middle row of Fig. 5 depicts the situation wherein an individual's response, depicted in cell **f**, might be greater than expected because it reflects a cumulative build-up of similar events depicted in cells **d** and **e**. The bottom row of Fig. 5 depicts the opposite situation, wherein an individual's response, depicted in cell **i**, might be less than expected because the event that might trigger this response, **g**, has been cancelled out or nullified by another event **h**.

Thus, using the properties of light as a metaphor one can model how patterns of thought and behavioral responses are made up of components, which can be addressed one by one.

We have only scratched the surface of the metaphorical representational system but that is the basic idea. (See [7] for further details of this representational system.) It aims to have both therapeutic value as a form of art therapy wherein the artistic outcomes have relatively straightforwardly interpretable symbolic meanings, and as a tool for doing psychological research for gaining insight into human nature.

### III. Implementation

Static images of light are of limited use as a metaphor for understanding real human situations, which are dynamic, and which could unfold different ways in different conditions. We now examine two complimentary technologies for realizing the dynamic potential of the metaphor, first in an immersive installation, and second in a piece of software.

#### A. The LIVEIA Installation

The LIVEIA project, still in its infancy, is an aesthetically appealing and inspirational interactive installation that will enable users to visualize and creatively experiment with light-based representations of themselves and others, thereby clarifying feelings and events going on in their lives. It is intended for use in museums, galleries, and educational settings. In the tradition of artists such as Dan Flavin, Bruce Naumen, or James Turrell, light is used not just to illuminate something else; it is an intrinsic part of the artistic conception. In the sense of light acting as a representation of a conscious being, the approach bears some relationship to Popat and Palmer's [8] work with people interacting and connecting with light sprites through dance. The project aims to exert a meaningful impact on how people explore and develop understanding of the ways people are interconnected, and of the relationship between life events and artistic ideas.

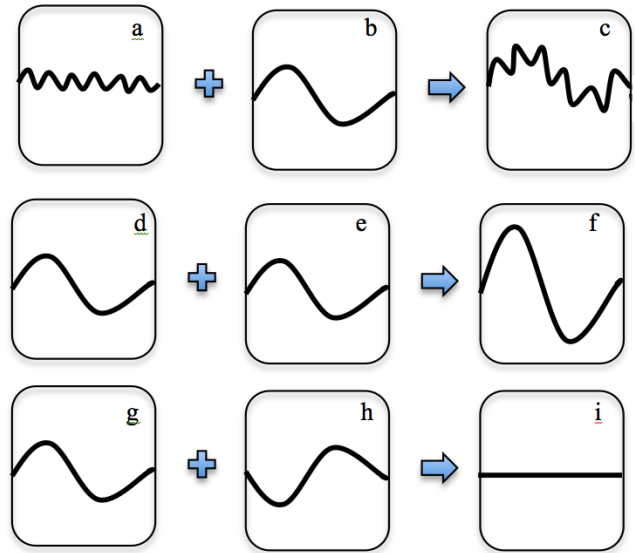

Figure 5. Top: Wave c is the sum of the superposed waves a and b. Centre: The amplitude of wave f is equal to the sum of its component waves d and e due to constructive interference. Bottom: Waves g and h exhibit destructive interference such that the result is no wave at i.

The allegory of Indra's Net (see above) will be written on a display at the entrance. Extending around the perimeter of a dimly lit, mirrored room will be a slowly undulating web of aluminum tubing covered with EL-wire, guided at the edges by rollers. At every vertex is a resin sphere containing a candle. Candles can be lit by attendees, and put into a spot from which they are drawn into the web. The goal of this first chamber of the installation is to remind attendees that we are all connected, and to have this be alive in their memories as they explore situations of potential isolation or conflict.

In the main room, a person's creative life force is represented as light, and their internal model of reality, or worldview, is portrayed as a spherical entity that amplifies and transforms this light. Attendees generate visual depictions of their inner workings by controlling how light moves through and between movable acrylic spheres of different sizes, colors, and degrees of transparency, hanging 4 to 6 feet above ground. They represent the hidden dynamics of their inner lives and relationships using handheld devices and toggles that control the properties of the spheres, including their relative positions and how they interact. Some spheres are detachable, allowing for physical, embodied interaction with them, which is expected to enrich attendees' experiences [9]. Some spheres are non-detachable because they are connected by servo motor links to projectors and a computer, and operated by toggles and handheld devices attached to control panels.

A toggle lets you vary the amount of light in a sphere to indicate how vibrant or alive the person is. Other toggles allow you to vary properties of the sphere such as the thickness, transparency, or colour of its outermost shell. The user is encouraged to use a thick shell to represent someone who is guarded and reserved, and a thin shell to represent someone who is open and friendly. There are toggles that allow the user to 'paint' different regions of the shell different colors, representing different arenas of life, or to create multiple spheres embedded in one another, i.e. 'layers of the self'.

Another set of toggles is used to represent thinking and communicating using beams of light. The user is invited to explore how varying the size, intensity, color, diffuseness, and direction of the beam relative to the sphere(s) affects reflection and refraction at sphere interfaces, and shown how these parameters can be used to represent phenomena such as reflection on an idea, miscommunication, deep versus superficial ideas, and so forth, as described in the previous section. Another toggle lets the user depict attraction or creative resonance between people as sparks that appear to flicker between the two spheres.

Yet another set of toggles enables the user to represent phenomena such as deception and repression, by generating fractures, dents, and internal boundaries of various shapes, sizes, and degrees of transparency, that distort how light flows in a sphere. The user can represent an emotionally charged topic by creating an opaque bubble in the sphere. The fact that light has difficulty both entering the bubble and leaving the bubble represents that the person probably either avoids the topic, or dwells on it excessively (or both, at different times). The user is invited to experiment with how fragmentation generates shadowy regions, and how the presence of shadowy regions leaves a detectable trace in the overall appearance of the sphere and the beams of light that leave it.

*B. SoulTracker Software*

The LIVEIA immersive environment is based on an earlier project, SoulTracker: an interactive virtual reality for depicting and playing with representations of inner light of self and others. It works along the same principles as the installation, and enables users to do everything they did there, and more. In addition to providing the capacity to model oneself and others using spheres and beams of light, it also provides:

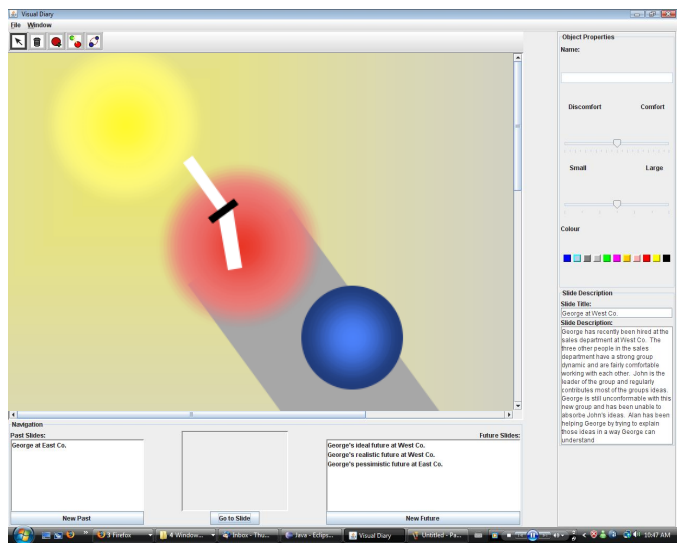

Figure 6. A screendump from SoulTracker (see text for details).

- More ways to depict aspects of one's inner life and interactions with others. For example, the degree of comfort people feel in each other's presence can be indicated by how blurred or sharp-edged their borders are when they are depicted in close proximity.
- The ability to make a copy of any scenario and use it as a starting point for depicting how it looks from different perspectives, how it arose, and what might happen next.
- The ability to create cross-sections of spheres and thereby see inside opaque spheres.
- The ability to save visualizations, email them, print them, post them to the cloud, etc.

Thus SoulTracker facilitates awareness and exploration of the potential each moment holds for creatively reconceptualizing the present and putting favorable future outcomes within reach. A working prototype has been built. The screendump in **Fig. 6** gives a simple example of how it is used to portray a hypothetical situation. Using *view mode* the user depicted two bonded individuals (fuzzy spheres), one of whom is deceiving the other (refracted beam), and a third (sharp-edged sphere in shadow) who is left out. In *overview mode* a miniature version of the situation would be shown, with events that led up to it to the left, and ways it could unfold to the right. The SoulTracker is still in a rudimentary state. Shadows and beams are not realistic, and the physics of light has not been built in; for example, reflection and refraction of light at surfaces and fractures must be drawn manually rather than happening automatically.

In the new version of SoulTracker, the physics of how light reflects and refracts off concave/convex surfaces and fractures, and how it is affected by structural features of spheres (e.g., diffuses, amplifies, or changes color) will be built in. This will make it easier to not just portray life situations but experiment with them and observe the consequences of modifying aspects of them. For example, wrestling with a problem will be represented as a beam of light that repeatedly reflects within the sphere, and to some degree refracts and escapes at the sphere's periphery, thereby subtly changes the ambient light and affecting others. Other features will be added, such as the ability to create the appearance of slowing down beams of light so that the user can watch the internal workings of spheres and their interactions. For example, it will be possible to observe the process by which a sphere regains equilibrium after receiving an incoming light beam (which represents assimilating new information), or direct light at fragmented regions until shadows disappear. The software will use a neural network to learn from user inputs to make suggestions for how depicted scenarios might unfold based on similar situations that have been entered before. It will be possible to view (and modify) how a given scenario might look from the perspective of someone else by clicking the sphere that represents that person. Members of families and organizations will be able to share their different perspectives of the same situation, and thereby learn from each other. By posting their scenarios to the cloud users will be able to locate others who are in similar situations to their own, and show each other how they are coping and feeling, i.e., communicate using this "language of light" even if they don't speak the same language.

## IV. Assessment of the Technology

Consenting users of the interactive installation will be filmed as they create and work with visual depictions of their inner processes. They will be interviewed about what insights they have gleaned about themselves, their relationships, their creative process, and their sense of purpose and self-understanding through this process. Exposure to the installation is expected to enhance users' understanding and control over their inner lives and interpersonal situations, and give evidence of breaking out of habitual patterns and approaching situations in new and more effective ways, as well as a sense of being part of something larger than oneself.

The effectiveness of SoulTracker as a therapeutic tool for artists will be investigated in experiments carried out over ten 90-minute sessions with consenting users. They will be divided into an experimental group that uses SoulTracker and a control group that watches entertaining videos. Both groups will be given a questionnaire to assess their degree of awareness and control over life problems and interactions with others, and capacity to achieve resolution and self-understanding through their creative process. Introduction to SoulTracker in the first session will be accompanied by discussion of the scientific and psychological framework. After that, and in subsequent sessions carried out at computers with a facilitator available to help and answer questions, they will be encouraged to depict and play with thoughts, creative ideas, situations and possible future developments of them.

SoulTracker sessions from consenting users will be anonymously analyzed for evidence of (1) sense of control over and understanding of situations, (2) awareness of the present moment and its rich possibilities for creative decision making and breaking out of habitual modes of thinking and acting, (3) ability to communicate, take into account perspectives of others and respond with empathy, and (4) evidence of enhanced understanding or resolution of life events through creative re-interpretation of them. In addition, the questionnaire given in the first session will be re-administered to both groups to assess the therapeutic value of the SoulTracker. A second questionnaire will be administered to the experimental group only to determine how personally useful they found the SoulTracker.

## V. Expected Results

Use of these tools is expected to facilitate (1) enhanced appreciation for the infinite number of ways of constructing the tapestries of understanding from which our thoughts and actions emanate and thereby affect others, (2) the weaving of difficult to verbalize but emotionally charged situations into a form in which it is possible to comprehend, explore, and come to terms with them, and (3) creative problem solving. By better understanding how users are using the software and how their use changes across sessions we can gain a better understanding of its effectiveness and how to improve it. Analysis of the actual scenarios generated and accompanying written comments using the SoulTracker sessions is expected to provide evidence of self-discovery and of new ways of using the software to shed light on human nature.

## VI. Collection and Application of Big Data

LIVEIA and SoulTracker will be made available over the internet and so useable by anyone in the world. Data from consenting users will be analyzed to determine how they are being used, and make them easier and more effective for users. One application of the data aims to enable like-minded individuals to gain access to one another. Individuals who develop similar models of themselves, or who tend to fall into similar life situations, will be able to contact one another if they are mutually agreeable to this. Even if they do not speak the same language it may be comforting to know that someone, somewhere in the world, is experiencing something similar to what oneself is experiencing.

Perhaps the most exciting application of the data is to build a "smart" version of LIVEIA. Sequences of visualizations that depict successful ways of working through difficult situations will be stored and when new individuals appear to be going through similar situations, the solutions found by their peers will be made available to them. For example, in a family dispute it may be helpful to re-orient the depictions of the various family members so that someone who was depicted on the periphery is now in the middle, to facilitate the ability to see the situation from that person's perspective. Actions such as this, when they prove helpful, will be stored in memory and retrieved for other users when appropriate, to provide suggestions for how to resolve situations.

## VII. Theoretical Framework and Next Steps

The work presented here grew out of earlier applications of optics to model the cultural evolution of human worldviews [10,11]. More broadly the research program arose as part of a scholarly effort to develop a scientific framework for cultural evolution based on the hypothesis that what evolves through culture is integrated internal models of the world, or worldviews: including stories, memories, knowledge, values, and beliefs, with both cognitive and emotional components [12,13,14]. In order for this research to exert a significant impact on how culture actually evolves, it is necessary to not just study cultural change from the outside but provide opportunities for growth on the inside. Such opportunities for growth must be broadly intuitive and engaging. LIVEIA provides opportunities to visualize ones' worldview as part of an evolving tapestry of interacting worldviews. It prompts micro-moments of reflection on one's ways of being and relating, and could therefore affect the myriad thoughts and acts that together constitute human cultural evolution.

The research program generates avenues for further investigation. A next step involves getting a better handle on what we mean by the term inner light. One means of accomplishing this involves assessing the extent to which there is agreement amongst people's assessments of the degree to which someone exudes (or obstructs) their inner light using a modified version of a research protocol that has been previously used to assessing the extent to which there is agreement amongst people's assessments of the degree of authenticity in creative performances [15]. Another avenue involves using the tool to better understand and track the creative process, and test theories about how it works [16].

## VIII. Summary and Conclusions

This paper described a new technology, the implementation of which is still in progress, that aims to have a meaningful impact on how people explore and develop ways of understanding themselves, their creative process, and their relationships to others and to their community. The work described here is at an early stage. We are currently seeking funds to work on the project and collaborators in computer graphics, optics, and digital technology. Our plans include not just building but testing the effectiveness of the technology.

There are two components, unified by an underlying metaphor between physical light and inner light, which can refer to creative spark, life force, or spiritual essence. Creative life force is portrayed as ambient light, and peoples' psyches are represented by spheres that amplify and transform light. Personality characteristics, situations, and relationships can be systematically depicted using a systematic visual language based on the properties of light and how it interacts with physical objects. For example, vibrant people are portrayed as having lots of light, and aloof people as having spheres with thick shells. Thoughts and ideas are represented as beams of light that converge—become more focused—when reflected off the concave inner surface of a sphere, and refract (bend) when they are (intentionally or unintentionally) misunderstood. Vague ideas are represented with diffuse beams that require much reflection. Thus the metaphor turns elusive aspects of human nature, and the situations we find ourselves in, into concrete visual depictions that can be explored and experimented with.

LIVEIA enables attendees to visualize and creatively experiment with light-based representations of themselves and others. Handheld devices attached to control panels enable attendees to direct beams of light of different sizes and intensities through and between spheres. Toggles enable attendees to control qualities of a sphere such as its color, opacity, and level of ambient light, and the user can generate regions of fragmentation that create shadows and distort the flow of light within a sphere. SoulTracker works along the same principles as LIVEIA, but offers a more private forum for visually depicting peoples' inner processes using spheres and beams of light, and provides some extra features. For example, the degree of comfort people feel in each other's presence can be indicated by how blurred or sharp-edged their borders are when they are in close proximity. It also allows users to save depicted scenarios, and to use them as a starting point for depicting how the same situation looks from different perspectives, or how it arose, or to create different possible 'future scenarios' for the depicted situation.

In short, new technologies are useful for visualizing and understanding psychological phenomena that underlie social wellbeing such as communication and miscommunication, wholeness and fragmentation, honesty and dishonesty, closeness and isolation, potentiality and actualization, and the process by which ideas are born and take shape. It is hoped providing a means to visually depict the intangible but all-important psychological and spiritual elements of life using a straightforward "language of light" will facilitate the evolution of worldviews that are integrated, adaptive, and humane.

Acknowledgments

This research is funded by a grant from the National Science and Engineering Research Council of Canada.